\documentclass[runningheads]{llncs}
\usepackage[T1]{fontenc}
\usepackage{enumitem}
\usepackage{graphicx,verbatim}
\usepackage{amsfonts}  
\usepackage{amsmath}
\usepackage{multirow}
\usepackage{booktabs}
\usepackage{xcolor}
\usepackage{colortbl}
\usepackage{array}
\definecolor{lightblue}{rgb}{0.85, 0.93, 1}
\usepackage{hyperref} 

\begin{document}
%
\title{Interactive Tumor Progression Modeling via Sketch-Based Image Editing}
%
%

\author{
    Gexin Huang\textsuperscript{1} \and 
    Ruinan Jin\textsuperscript{1} \and 
    Yucheng Tang\textsuperscript{2} \and 
    Can Zhao\textsuperscript{2} \and 
    Tatsuya Harada\textsuperscript{3,4} \and 
    Xiaoxiao Li\textsuperscript{1} \and 
    Gu Lin\textsuperscript{3,4}
}

\authorrunning{G. Huang et al.}

\institute{
    \textsuperscript{1} University of British Columbia, Canada \\
    \textsuperscript{2} NVIDIA, USA \\
    \textsuperscript{3} University of Tokyo, Japan \\
    \textsuperscript{4} RIKEN, Japan \\
    \email{gexinh@student.ubc.ca, ruinanjin@alumni.ubc.ca, 	yucheng.tang@vanderbilt.edu, 	canz@nvidia.com, harada@mi.t.u-tokyo.ac.jp, xiaoxiaoo.li@ece.ubc.ca, lin.gu@riken.jp}
}
    
%
%
\maketitle              
\begin{abstract}
Accurately visualizing and editing tumor progression in medical imaging is crucial for diagnosis, treatment planning, and clinical communication. To address the challenges of subjectivity and limited precision in existing methods, we propose SkEditTumor, a sketch-based diffusion model for controllable tumor progression editing. By leveraging sketches as structural priors, our method enables precise modifications of tumor regions while maintaining structural integrity and visual realism. We evaluate SkEditTumor on four public datasets—BraTS, LiTS, KiTS, and MSD-Pancreas—covering diverse organs and imaging modalities. Experimental results demonstrate that our method outperforms state-of-the-art baselines, achieving superior image fidelity and segmentation accuracy. Our contributions include a novel integration of sketches with diffusion models for medical image editing, fine-grained control over tumor progression visualization, and extensive validation across multiple datasets, setting a new benchmark in the field. The code is available at \href{https://anonymous.4open.science/r/SketEdit-E0DC}{here}.

\end{abstract}

\section{Introduction}

\begin{figure}[ht]
	\centering
	\includegraphics[width=\linewidth]{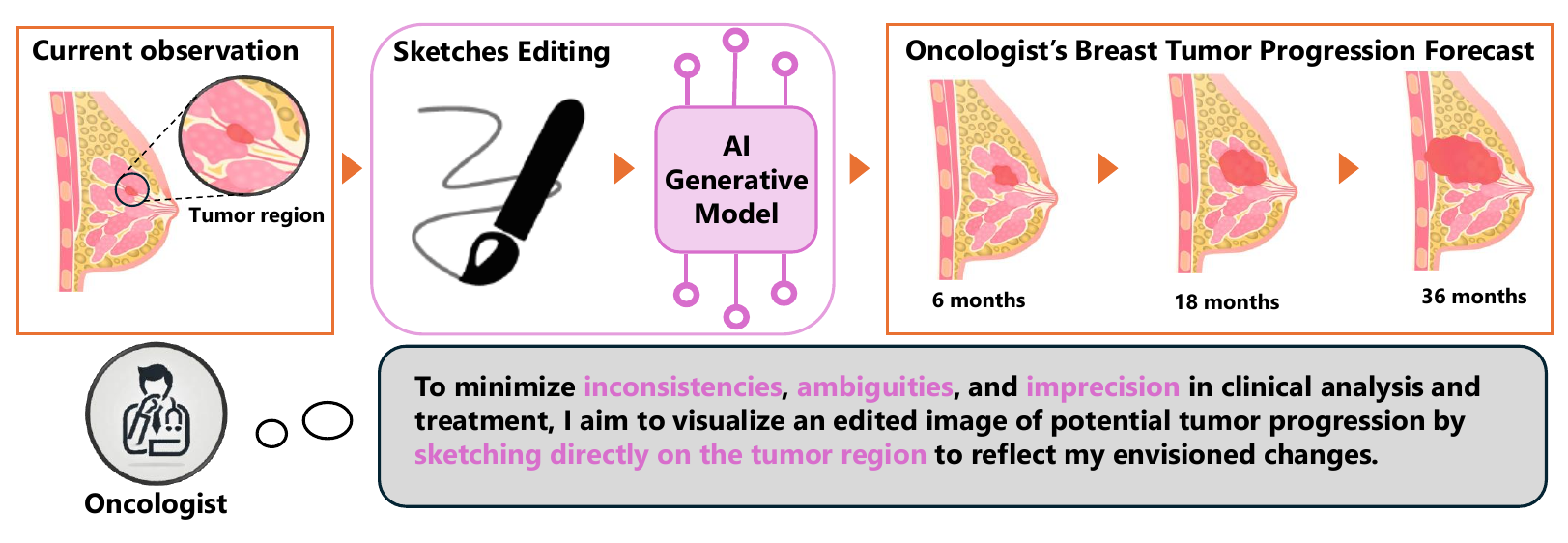}
	\caption{Illustration of the proposed sketch-guided tumor progression modeling approach. The user sketches directly on the tumor region to generate an edited image, simulating potential tumor growth or regression based on the envisioned changes.}
	\label{fig_1}
     \vspace{-10pt}
\end{figure}

With rapid advancements in non-invasive imaging technologies, medical imaging has become critical for identifying, characterizing, and managing tumors by providing key insights into tumor location, volume, and metastatic spread. However, monitoring subtle tumor changes remains challenging due to subjective image interpretation~\cite{o2017imaging}, variable tumor presentations~\cite{lambin2012radiomics}, and limited longitudinal data~\cite{yip2016applications}, especially in prostate cancer~\cite{stabile2020multiparametric}. These issues can lead to inconsistencies, ambiguities, and reduced precision in clinical collaboration. As shown in the Fig.~\ref{fig_1}, sketch-guided image editing addresses these hurdles by enabling interactive, fine-grained modifications through sketch-based annotations, allowing clinicians to directly incorporate structural details and achieve more accurate visualizations of tumor evolution over time. This enhanced clarity fosters better communication among experts, minimizes misinterpretation, and improves patient understanding of disease progression and treatment options, ultimately strengthening clinical decision-making and outcomes.

To mitigate this, some studies have proposed using synthesized sketches, such as edge maps or semantic contours (i.e., boundaries of ground-truth semantic masks), as proxies for actual sketches during training. Zhang et al. \cite{zhang2019skrgan} introduced a sketch detection module that extracts sketches from edge maps, enabling GAN-based models to generate images across modalities. Liang et al. \cite{liang2020synthesis} leveraged edge maps and labels in a PatchGAN model to progressively refine ultrasound image generation. Wang et al. \cite{wang2021realistic} designed masks combining foreground and background in CT lung images to condition GANs for nodule and tissue synthesis, while Toda et al. \cite{toda2022lung} enhanced StyleGAN to generate diverse cancer CT images by controlling tumor shapes via Canny-detected edges. Despite these efforts, most methods focus on image generation rather than editing, offering limited user control and leading to inconsistent outputs for identical inputs. Although Fernando et al.\cite{perez2024radedit} and Alaya et al.\cite{alaya2024mededit} recently introduced conditional diffusion models for medical image editing, they use masks as conditioning inputs rather than sketches, limiting their applicability for fine-grained structural modifications. In contrast, sketch-based editing has been explored for human faces and natural images \cite{portenier2018faceshop,jo2019sc,liu2021deflocnet,zeng2022sketchedit,xu2023draw2edit}, but these methods heavily depend on large domain-specific datasets and are not readily transferable to medical imaging. Moreover, many of these approaches are GAN-based, requiring extensive hyperparameter tuning to mitigate mode collapse\cite{srivastava2017veegan}, often at the cost of quality and diversity in generated images\cite{chen2016infogan}.

To overcome these limitations, we propose SkEditTumor, a sketch-controlled tumor progression editing framework leveraging diffusion models. Diffusion models provide computational efficiency and robust performance across diverse datasets and imaging modalities, making them well-suited for medical imaging applications. Our approach incorporates a sketch encoder to embed annotations into the latent space, enabling precise control over tumor progression editing. This ensures realistic and consistent tumor evolution visualizations across various imaging modalities. Our key contributions are as follows: 1) We introduce the first sketch-guided image editing framework for modeling tumor progression in medical images; 2) By embedding sketch annotations into the latent representation, our method enables fine-grained control over tumor dynamics; 3) We validate our approach across multiple tumor datasets and imaging modalities, achieving superior visual quality and accuracy compared to baseline methods.

\begin{figure}[ht]
	\centering
	\includegraphics[width=\linewidth]{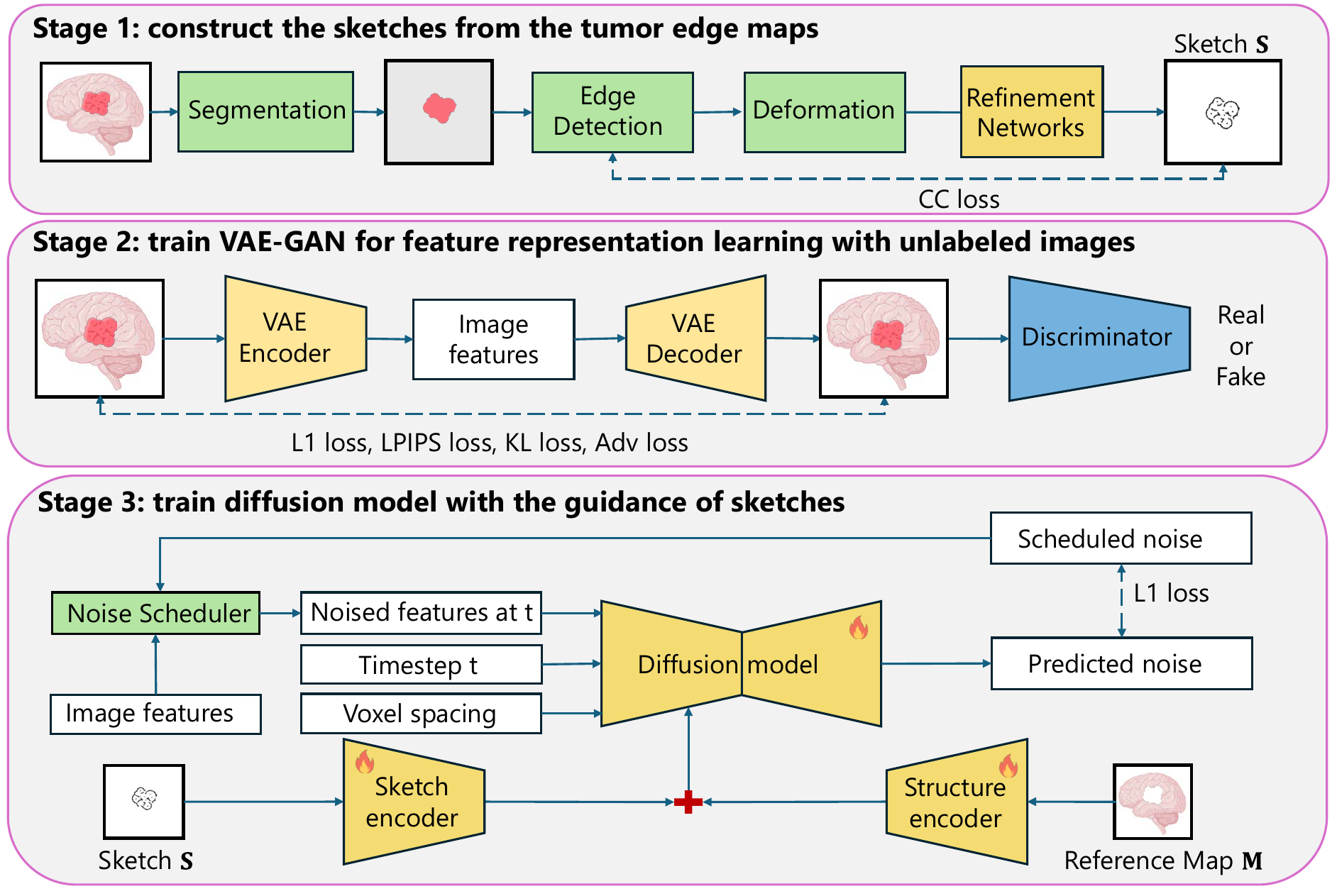}
	\vspace{-0.8cm}
	\caption{Overflow of the SkEditTumor framework. In Stage 1, the tumor region is segmented, and edge detection is applied to generate edge maps. A deformation module introduces variability to simulate hand-drawn sketch imperfections, followed by a refinement network to produce a precise sketch $\mathbf{S}$. In Stage 2, a VAE-GAN is trained to encode image features with a discriminator ensures realism through adversarial training. In Stage 3, a diffusion model integrates $\mathbf{S}$ and reference map $\mathbf{M}$ into its latent space, progressively denoising it to produce realistic and controllable tumor progression edits.
    }
	\label{fig_2}
     \vspace{-10pt}
\end{figure}

\section{SkEditTumor}

\subsection{Sketch Refinement Strategy}
Collecting medical images and their corresponding free-hand sketches is a challenging task. Early research for sketch refinement\cite{eitz2012humans,vinker2022clipasso} have proven that sketches are closely related to edges, both of which are visually closed outlines of objects. Thus, we aim to utilize the edge map of tumor region to form the corresponding sketch. However, the major difference between sketches and edges is that edges are pixel-wise corresponding to sharp intensity gradients, while sketches are more diversified and abstract. It means that the model trained with the edge map need the pixel-wise preciseness from the user-provided sketches and can hardly handle the real free-hand sketches.To address this, we propose a sketch refinement strategy.
Given an input image $\mathbf{X}$, we first obtain the segmentation mask $\mathcal{M}$ using Swin UNETR~\cite{hatamizadeh2021swin} and nnU-NET~\cite{isensee2021nnu}, followed by Canny edge detection to extract the binary edge map $\mathbf{E}$. To mimic real sketches, we apply random erosion $f_{ero}(\cdot)$ or dilation $f_{dia}(\cdot)$ (kernel size 3), generating an altered edge map $\mathbf{E^*}$ with structural variations. Next, we introduce deformation by defining a random displacement field $\mathbf{D}(\mathbf{E}^*)$, where each component \( de_i(\mathbf{E}^*) \) is sampled from a normal distribution \( \mathcal{N}(0, \sigma_0) \), with \( \sigma_0 \) controlling noise strength. A Gaussian filter smooths the displacement: $de_i'(\mathbf{E}^*) = \text{GaussianFilter}(de_i(\mathbf{E}^*))$. Finally, bilinear interpolation $f_{bi}(\cdot)$ maps pixels onto their displaced positions, producing an elastically deformed sketch: $\mathbf{S}^* = f_{bi}(\mathbf{E}^*, \mathbf{D}'(\mathbf{E}^*)).$ To refine $\mathbf{S}^*$ into a structured sketch while eliminating gaps, we train a U-Net refinement network $f_{u}$, yielding $ \mathbf{S} = f_{u}(\mathbf{S}^*).$ This approach ensures robustness to sketch imperfections, improving model adaptability to real user-provided sketches.

To enhance sketch refinement, we move beyond pixel-wise losses (L1/L2), which minimize pixel-level discrepancies but fail to capture global structure and regional correlations essential for free-hand sketches. These losses assume strict spatial alignment and overlook inter-region dependencies, limiting coherence in sketch refinement. To address this, we propose a region-wise cross-correlation (CC) loss that leverages mutual information between regions for more effective sketch refinement. Specifically, we calculate the local mean of pixel values within a sliding grid $p_i$ of size $N \times N$ as $\mathbf{S}^r_{p_i} = \frac{1}{N^2} \sum_{j=1}^N \sum_{i=1}^N v_{i,j}$, where $\mathbf{S}^r_{p_i}$ represents the mean refined sketch values within grid $p_i$, and $v_{i,j}$ denotes the the pixel intensity at position $(i,j)$. Similarly, the local mean for the edg map $\mathbf{E}$ in $p_i$ is $\mathbf{E}^r_{p_i}$.
The region-wise CC loss is formulated as the cross-correlation between the local means of these regions, normalized by their standard deviations, which is defined as:
\begin{equation}
\mathcal{L}_{CC}(\mathbf{S}, \mathbf{E}) = 
- \sum_{p \in P}
\frac{
\sum_{p_i \in p} \big(\mathbf{S}^r_{p_i} - \overline{\mathbf{S}^r}\big) \big(\mathbf{E}_{p_i}^r - \overline{\mathbf{E}^r}\big)
}{
\sqrt{\sum_{p_i \in p} \big(\mathbf{S}^r_{p_i} - \overline{\mathbf{S}^r}\big)^2 
      \sum_{p_i \in p} \big(\mathbf{E}^r_{p_i} - \overline{\mathbf{E}^r}\big)^2}
}.
\end{equation}
where $p$ is one of the region in the whole image $P$, and $\overline{\mathbf{S}^r}$ and $\overline{\mathbf{E}^r}$ are the global means of the pixel values for the region.

\subsection{Variational Autoencoder}
Diffusion models directly applied to medical image incur significant computational costs. To address this, Latent Diffusion Models (LDMs)~\cite{rombach2022high} operate within a compressed, lower-dimensional latent space. Thus, we built the the Variational Autoencoder (VAE) upon previous studies [19, 49] with the constraint of combined objectives, which integrates perceptual loss $\mathcal{L}_{lpips}$ [71], adversarial loss $\mathcal{L}_{adv}$ [68], and $L_1$ reconstruction loss $\mathcal{L}_{recon}$. These combined objectives ensure that the VAE adhere closely to the image manifold and enforce local realism. In addition, we follow \cite{kingma2013auto,rezende2014stochastic,rombach2022high} adding Kullback-Leibler (KL) regularization $\mathcal{L}_{reg}$ toward a standard normal on the learned latent features for avoiding high-variance latent spaces. Additionally, we incorporate data augmentation and patch cropping to enhance image details, following~\cite{guo2024maisi}. Given a medical image $\mathbf{X}$, the encoder $f_{enc}(\cdot)$ of AE downsamples $\mathbf{X}$ and generates the latent representation $\mathbf{Z} = f_{enc}(\mathbf{X}) \in \mathcal{R}^{h \times w \times d}$ with much smaller spatial dimensions. The decoder $f_{dec}(\cdot)$ of AE approximates the reconstructed image $\tilde{\mathbf{X}} = f_{dec}(\mathbf{Z})$ from the latent features. A discriminator, denoted as $\mathcal{D}$, is utilized to identify and penalize any unrealistic artifacts in the reconstructed volume $\tilde{X}$. After training the VAE, we utilize the latent feature $\mathbf{Z}$ as the input of the sketch-guided diffusion model.

\subsection{Sketch-guided Latent diffusion model}

We focus on editing tumor progression in medical images while preserving organ textures outside tumor regions. To achieve this, the latent diffusion model (LDM) is conditioned on two inputs: (1) a sketch $\mathbf{S}$, defining the tumor's shape and location in latent space, and (2) a reference map $\mathbf{M}$, providing contextual information to maintain the natural appearance of unaffected areas. Formally, LDM operates in a latent feature space rather than at the pixel level, transforming data between a simple noise distribution and the learned data distribution through a forward and reverse process. In the \textit{forward process}, the latent variable $\mathbf{Z}$ is progressively corrupted by Gaussian noise over $T$ time steps, converting the original data representation into white noise. At each step $t \in \{1, \dots, T\}$, the noisy latent variable is defined as: $\mathbf{Z}_t = \sqrt{\alpha_t} \mathbf{Z} + \sqrt{1 - \alpha_t} \epsilon, \quad \epsilon \sim \mathcal{N}(0, \mathbf{I})$, where $\alpha_t$ controls noise variance. By step $T$, the latent variable approximates a standard normal distribution: $\mathbf{Z}_{T} \sim \mathcal{N}(0, \mathbf{I})$. The \textit{reverse process} reconstructs $\mathbf{Z}$ from $\mathbf{Z}_{T}$ using a neural network $\epsilon_{\theta}(\cdot)$, which predicts and removes noise at each step. Starting from $\mathbf{Z}_{T}$, it iteratively refines: $\mathbf{Z}_{t-1} = f_{\theta}(\mathbf{Z}_t, t)$, where $f_{\theta}$, parameterized by $\epsilon_{\theta}$, is a time-conditional U-Net~\cite{ronneberger2015u} designed for hierarchical feature extraction and multi-scale noise handling.

To incorporate the sketch and reference map, we design a \textit{sketch encoder} and a \textit{structure encoder} to encode their respective information as conditions for the diffusion model, enabling sketch-based editing. The \textit{sketch encoder}, following~\cite{konz2024anatomically}, uses a compact convolutional network to transform the sketch $\mathbf{S}$ into latent features $\mathcal{C}_s$, aligning it with the latent space. Inspired by ControlNet~\cite{zhang2023adding}, trainable copies of the neural network blocks of $\epsilon(\cdot)$ are connected to $\epsilon_{\theta}$ via zero convolution layers, evolving from zero weights to optimal settings during training. For the reference map, we first generate a binary mask $\mathcal{M}^*$ from the sketch. Morphological operations ensure boundary continuity and close small gaps. A flood fill operation propagates through the background from a seed point outside the target region, followed by inversion to segment the sketch's interior. The reference map is computed as $\mathbf{M} = \mathcal{M}^* \odot \mathbf{X}$, where $\odot$ denotes the Hadamard product. The \textit{structure encoder} then extracts latent features $\mathcal{C}_r$ from $\mathbf{M}$ via a compact convolutional network, processed through trainable neural network blocks of $\epsilon(\cdot)$. The resulting intermediate features are combined with those from the sketch. Following~\cite{guo2024maisi}, the LDM is conditioned on voxel spacing instead of text prompts, using a vector $\mathbf{v}$ that encodes the physical size of each voxel along three dimensions. Unlike conventional frozen LDMs, we jointly train the LDM with the sketch and structure encoders, making it more suitable for editing tasks. The overall learning objective is $\mathbb{E}_{\epsilon \sim \mathcal{N}(0, 1), t, \mathbf{v}, \mathcal{C}_s, \mathcal{C}_r} \left[ \| \epsilon - \epsilon_{\theta} (\mathbf{Z}_t, t, \mathbf{v}, \mathcal{C}_s, \mathcal{C}_r) \|_1 \right].$ During inference, a reference image is selected, and a free-hand sketch $\mathbf{S}^*$ is drawn to represent tumor progression. The refinement network $f_{u}$ processes $\mathbf{S}^*$ into $\mathbf{S}$, which generates the binary mask $\mathcal{M}^*$ and reference map $\mathbf{M}$. Given the voxel spacing, the LDM reconstructs the edited latent features $\mathbf{Z}_{\text{tar}}$, which are decoded into the final image $\mathbf{X}_{\text{tar}} = \mathcal{D}(\mathbf{Z}_{\text{tar}})$.

\begin{figure}[t]
	\centering
	\includegraphics[width=\linewidth]{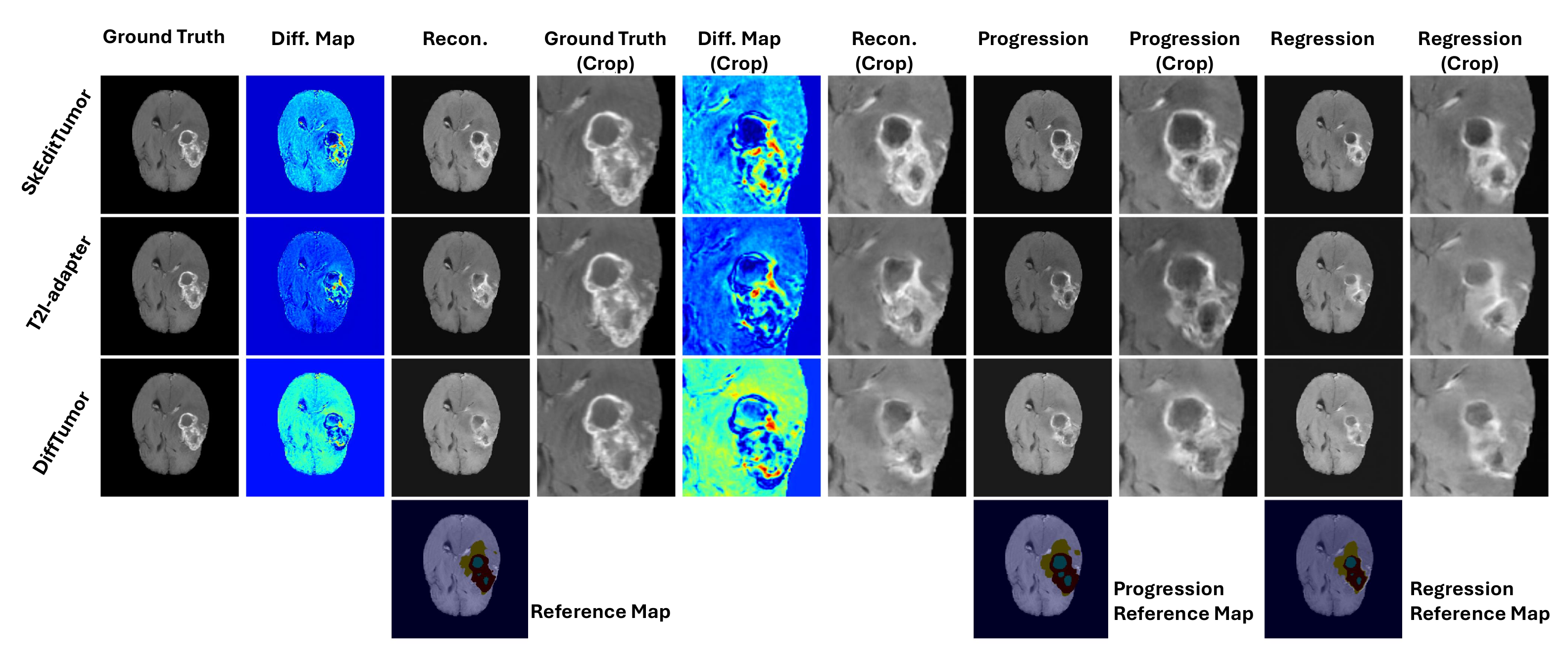}
	\vspace{-0.8cm}
	\caption{Reconstruction and editing results on the BraTS dataset. Qualitative comparison of SkEditTumor (ours), T2I-Adapter, and DiffTumor. Columns 1-3 show reconstruction results with unchanged sketches, including the ground truth, difference maps, and reconstructed images. Columns 4-6 display cropped tumor regions. Columns 7-12 demonstrate tumor progression and regression edits achieved by modifying the sketches, with SkEditTumor delivering more accurate and visually consistent results compared to baselines.}
	\label{fig_3}
     \vspace{-10pt}
\end{figure}

\begin{figure}[t]
	\centering
	\includegraphics[width=\linewidth]{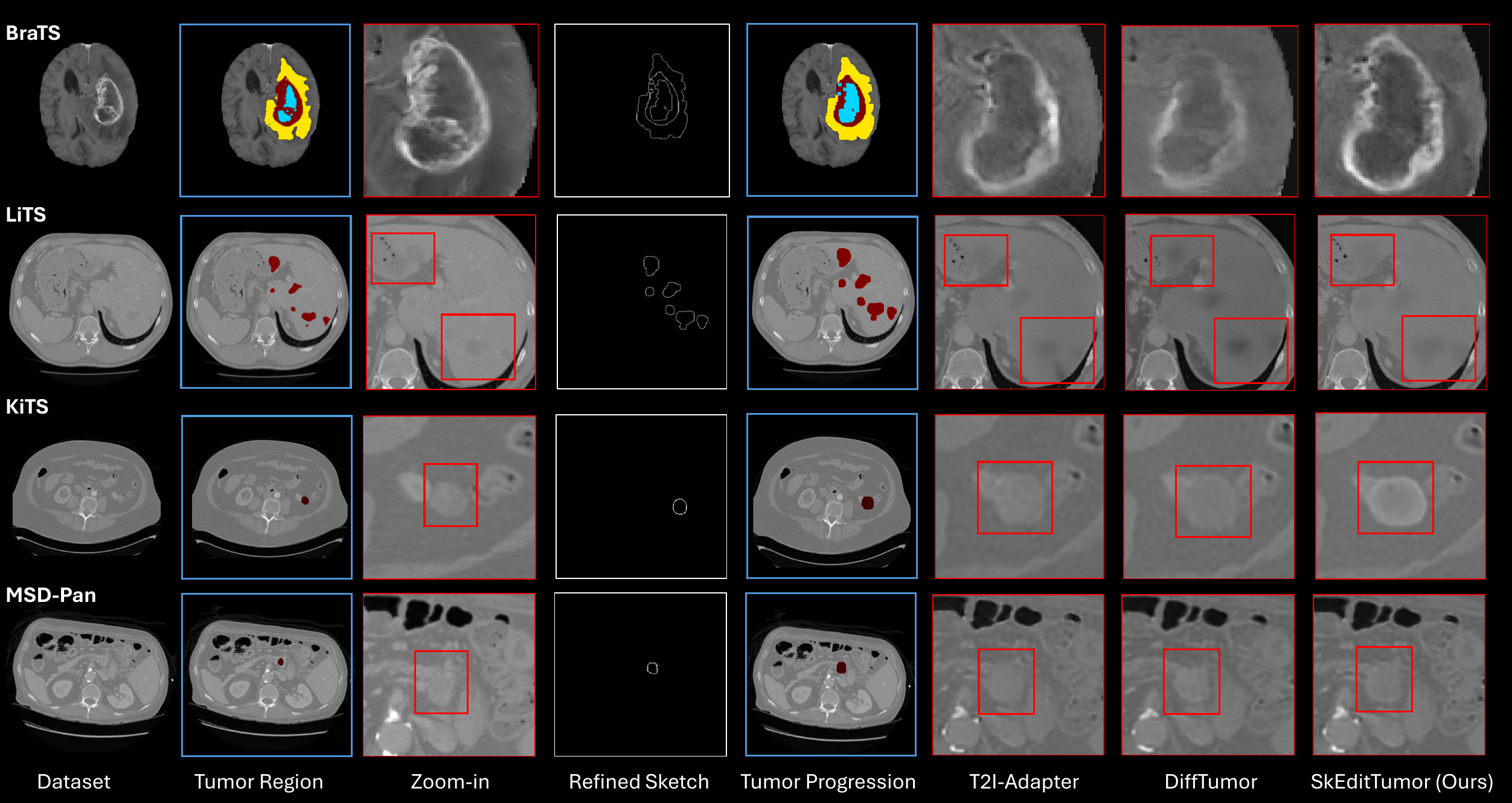}
	\vspace{-0.8cm}
	\caption{Tumor editing results across four datasets (BraTS, LiTS, KiTS, MSD-Pancreas). For each dataset, the input image, real tumor segmentation, and zoom-in view of the tumor region are shown, followed by the refined sketch and the corresponding segmentation for tumor expansion. The tumor editing results generated by T2I-Adapter, DiffTumor, and SkEditTumor (ours) are presented in the final three columns. The tumor regions (red boxes) highlight the differences in tumor structure preservation, with SkEditTumor producing more realistic and accurate edits compared to baseline methods.}
	\label{fig_4}
     \vspace{-10pt}
\end{figure}

\section{Experiments $\&$ Results}

\subsection{Datasets.} 
We evaluate our method on four publicly available datasets: BraTS~\cite{menze2014multimodal}, LiTS~\cite{bilic2023liver}, KiTS~\cite{heller2019kits19}, MSD-Pancreas~\cite{antonelli2022medical}, and PI-CAI~\cite{saha2022pi}, covering diverse organs, tumor types, and imaging modalities. The \textbf{BraTS} dataset includes 469 multi-modal MRI scans (T1, T1ce, T2, FLAIR) with glioma annotations. We use T1ce due to its superior tumor contrast. The \textbf{LiTS} dataset contains 131 contrast-enhanced CT volumes with liver tumor annotations, challenging due to heterogeneous intensities. The \textbf{KiTS} dataset provides 489 CT scans with kidney tumor and renal parenchyma segmentations, covering tumors of varying size and morphology. The \textbf{MSD-Pancreas} dataset consists of 201 CT volumes with pancreas and pancreatic tumor annotations, characterized by low contrast and irregular shapes. 

All datasets are preprocessed using MONAI~\cite{cardoso2022monai}, with intensity normalization and uniform voxel spacing. Intensities are scaled to [0,1] with modality-specific adjustments: MRI values are clipped to the 0th–99.5th percentile before scaling, while CT values are clipped to [-1000, 1000] HU. These steps ensure consistency across organs, tumor types, and imaging modalities for fair comparisons. Finally, dataset are split into the training set and the test set as the ratio of 8:2. 


\subsection{Experimental Setting and Evaluation Metrics}
We use the Adam optimizer with $\beta_1 = 0.9$ and $\beta_2 = 0.999$. The autoencoder (AE) is trained with a batch size of 64 for 20,000 steps at a learning rate of $1 \times 10^{-4}$, selected to prevent image blurring from KL divergence regularization. Larger batch sizes significantly improved training outcomes. For both datasets, diffusion models were trained with a batch size of 20 for 40,000 steps using the same learning rate $1 \times 10^{-5}$. We employ the state-of-the-art segmentation model nnU-NET~\cite{isensee2021nnu} for evaluating the segmentation accuracy of edited images. To assess SkEditTumor, we compare it against two baselines: T2I-Adapter~\cite{mou2024t2i}, which integrates text-to-image generation with structural priors for guided editing, using the sketch and reference map as conditions; and DiffTumor~\cite{chen2024towards}, a diffusion-based tumor synthesis method where we provide the sketch and reference as inputs. All experiments were conducted on a single NVIDIA A100 GPU.

To evaluate our framework and baselines, we employ metrics for image quality and segmentation accuracy. NRMSE evaluates pixel-level deviations, with lower values reflecting higher accuracy. SSIM measures structural similarity, luminance, and contrast, with higher values indicating better preservation of structural integrity. PSNR quantifies reconstruction fidelity by assessing the ratio of signal power to noise, where higher values indicate less distortion. For tumor region precision, we compute the Dice score, measuring overlap between predicted and ground truth tumor regions, providing critical insights into medical relevance.

\begin{table}[!htbp]
\centering
\caption{Quantitative assessment of image generation fidelity across four datasets (BraTS, LiTS, KiTS, and MSD-Pancreas). We compare T2I-Adapter and DiffTumor with our SkEditTumor framework under three condition settings: (1) “+” indicates using accurate edges as the condition, (2) no “+” indicates using the refined sketches, and (3) “w/o R” denotes using free-hand sketches (no refinement network). We report NRMSE↓, SSIM↑, PSNR↑, and Dice↑. Blue cells highlight the best performance for each metric and dataset.}
\label{tab1}
\renewcommand{\arraystretch}{1.05}
\setlength{\tabcolsep}{2pt}

\begin{tabular}{c|cccc|cccc}
\toprule
\multirow{2}{*}{Method} & \multicolumn{4}{c|}{BraTS} & \multicolumn{4}{c}{LiTS} \\ \cline{2-9} 
 & NRMSE↓ & SSIM↑ & PSNR↑ & Dice↑ & NRMSE↓ & SSIM↑ & PSNR↑  & Dice↑\\ \hline
T2I-adapter+~\cite{mou2024t2i}    & 0.040  & \cellcolor{lightblue}\textbf{0.918} & 31.54 & 81.9 &            0.109 & 0.793 & 26.60  & 58.1\\
DiffTumor+~\cite{chen2024towards} & 0.043  & 0.915 & 30.61 & 80.3 &            0.111 & \cellcolor{lightblue}\textbf{0.798} & 26.54  & 57.2 \\
SkEditTumor+                      & \cellcolor{lightblue}\textbf{0.035}  & \cellcolor{lightblue}\textbf{0.918} & \cellcolor{lightblue}\textbf{31.47}  & \cellcolor{lightblue}\textbf{82.4} &            \cellcolor{lightblue}\textbf{0.107} & 0.788 & \cellcolor{lightblue}\textbf{26.72}  & \cellcolor{lightblue}\textbf{60.2}\\   
SkEditTumor                       & 0.048  & 0.910 & 29.07 & 78.3 &            0.113 & 0.710 &  25.88 &  57.1\\ 
    w/o R                             & 0.051  & 0.900 & 28.55 & 74.5 &            0.138 & 0.605 & 22.20 &  54.2\\

\bottomrule
\toprule

\multirow{2}{*}{Method} & \multicolumn{4}{c|}{KiTS} & \multicolumn{4}{c}{MSD-Pancreas} \\ \cline{2-9} 
 & NRMSE↓ & SSIM↑ & PSNR↑  & Dice↑ & NRMSE↓ & SSIM↑ & PSNR↑  & Dice↑ \\ \hline
T2I-adapter+~\cite{mou2024t2i}     & 0.028 & 0.933 &  31.88 & 73.7 &        0.054 &  0.732  &  29.00   &  67.3 \\
DiffTumor+~\cite{chen2024towards}  &  0.030 &  0.927  & 31.08 &  72.3 &      0.083 &  0.695  &  28.50  &  66.4 \\
SkEditTumor+  & \cellcolor{lightblue}\textbf{0.024} & \cellcolor{lightblue}\textbf{0.942} &  \cellcolor{lightblue}\textbf{32.59} & \cellcolor{lightblue}\textbf{74.1}                & \cellcolor{lightblue}\textbf{0.026} &  \cellcolor{lightblue}\textbf{0.857} &  \cellcolor{lightblue}\textbf{31.75}   &  \cellcolor{lightblue}\textbf{69.0}   \\
SkEditTumor   &  0.031 & 0.915   &  30.61  & 72.0          & 0.031 & 0.797 & 30.34  &  66.7    \\
w/o R         & 0.112 & 0.803 &  19.03 & 65.2                & 0.121 & 0.692  &  18.32   &  48.1   \\

\bottomrule
\end{tabular}%
\end{table}
\subsection{Qualitative Analysis}


We conducted a qualitative assessment of tumor editing—including reconstruction, progression, and regression—on the BraTS, LiTS, KiTS, and MSD-Pancreas datasets using T2I-Adapter, DiffTumor, and our proposed SkEditTumor framework. As illustrated in Fig.~\ref{fig_4}, SkEditTumor consistently demonstrates superior visual fidelity and morphological coherence across all datasets. In BraTS, T2I-Adapter and DiffTumor frequently produce tumor outlines that are either ambiguous or poorly aligned with the intended regions. In LiTS, DiffTumor’s edited tumors often display abnormally low HU intensities, diminishing radiological realism and pathophysiological plausibility. By contrast, SkEditTumor preserves anatomically relevant boundaries and yields more realistic intensities, effectively capturing subtle lesion characteristics. This advantage is especially evident in KiTS and MSD-Pancreas, where SkEditTumor produces clearer tumor margins and more distinct intratumoral structures. In comparison, T2I-Adapter and DiffTumor struggle with tumor expansion, resulting in blurred edges and reduced internal clarity. Overall, these findings underscore SkEditTumor’s robust editing capabilities and its effectiveness in delivering high-quality, anatomically coherent tumor modifications across diverse datasets.



\subsection{Quantitative Analysis}
To quantitatively assess fidelity, Table ~\ref{tab1} reports NRMSE, SSIM, PSNR, and Dice scores for all methods. SkEditTumor consistently demonstrates either superior or closely competitive performance. On LiTS, SkEditTumor+ attains the lowest NRMSE (0.107), indicating more realistic intensity reconstruction, while on KiTS it achieves the highest SSIM (0.942) and PSNR (32.59), reflecting sharper boundaries and reduced image distortion. Removing the refinement module (“w/o R”) notably degrades performance (e.g., SSIM drops from 0.942 to 0.893 on KiTS), confirming the pivotal role of the refinement network in maintaining anatomical plausibility. Nevertheless, using refined sketches (SkEditTumor) only slightly trails the variant that relies on perfectly accurate edges (SkEditTumor+), demonstrating robust performance even with less precise user inputs. Collectively, these results underscore SkEditTumor’s capacity to deliver high-fidelity tumor edits across diverse datasets, preserving both visual realism and structural consistency under varied sketching conditions. These results confirm that images generated by SkEditTumor not only exhibit superior visual fidelity but also retain structural consistency, enabling higher segmentation accuracy in downstream tasks. 



\section{Conclusion}

We propose a sketch-based diffusion model for tumor progression editing in medical imaging, leveraging sketches as a prior within a conditional diffusion pipeline for precise, controllable modifications. Validated on brain MRI and CT tumor progression tasks, our method achieves state-of-the-art performance both quantitatively and qualitatively. Clinically, it enhances communication between clinicians and patients, improves visualization of tumor dynamics, and supports treatment planning through realistic, interpretable edits. Currently limited to 2D images, the model requires further optimization and computational efficiency improvements for 3D extensions, which will be addressed in future work to ensure broader clinical applicability.
%
%
%
\bibliographystyle{splncs04}
\bibliography{mybibliography}

\end{document}